\documentclass[3p,times,procedia]{elsarticle}
\usepackage{nupha_ecrc}

\volume{00}

\firstpage{1}

\journalname{Nuclear Physics A}

\runauth{
}

\jid{nupha}

\jnltitlelogo{Nuclear Physics A}

\usepackage{amssymb}
\usepackage{lineno}

\usepackage[figuresright]{rotating}

\begin{document}

\begin{frontmatter}



\dochead{}
%% e.g. \dochead{17th International Conference on Dynamical Processes in Excited States of Solids}

\title{ Remembrance of things past: non-equilibrium effects and the evolution of critical fluctuations near the QCD critical point}


 \author[1]{Swagato Mukherjee\footnote{Presenter}}
 \author[1,2]{Raju~Venugopalan}
 \address[1]{Physics Department, Brookhaven National Laboratory, Upton, NY 11973, USA}
\address[2]{Institut f\"{u}r Theoretische Physik, Universit\"{a}t Heidelberg, Philosophenweg 16, 69120 Heidelberg, Germany}

 \author[1]{Yi Yin}


%\address{}

\begin{abstract} 
We report on recent progress in the study of the evolution of non-Gaussian cumulants of critical fluctuations. 
We explore the implications of non-equilibrium effects on the search for the QCD critical point. 
\end{abstract}


\end{frontmatter}

% \linenumbers

\section{Introduction}
\label{sec:intro}
The search for the conjectured QCD critical point has attracted much theoretical and experimental effort. 
%One universal feature of a system near the critical point  is the growth and divergence of the equilibrium correlation $\xi_{\rm eq}$ of critical modes. 
The fluctuations of critical modes scale with 
the equilibrium correlation length 
$\xi_{\rm eq}$ and consequently become prominent near the critical point.
%Since critical field $\sigma$ is coupled to the net baryon number density. 
%The contributions to the moments of net baryon number fluctuations are expected to be proportional to the corresponding moments of the critical field itself.
Since critical fluctuations contribute to the moments of net baryon number fluctuations, 
these fluctuation observables are considered to be tell tale signs of the presence of the QCD critical point~\cite{Stephanov:1999zu,Asakawa:2015ybt}. 

Of particular importance are the non-Gaussian skewness and kurtosis cumulants, which are respectively the third and fourth cumulants. 
They are more sensitive to the growth of the correlation length $\xi_{\rm eq}$ compared with the Gaussian cumulants~\cite{Stephanov:2008qz}. 
Furthermore, even qualitative features of the non-Gaussian cumulants such as a change in sign, and the associated non-monotonicity, 
can signal the presence of criticality in the QCD phase diagram~\cite{Stephanov:2011pb}. 
The beam energy dependence of those non-Gaussian cumulants have been measured by the STAR experiment; these are discussed in some of the  contributions to these proceedings. 

However, nature is secretive because the QCD critical point is shielded by non-equilibrium effects: 
developing a larger correlation length requires a longer time. 
%On the theoretical side, 
%a crucial challenging is from 
Universality arguments~\cite{Hohenberg:1977ym} indicate that the relaxation time of critical modes 
%$\tau_{\rm eff}\sim \xi^{z}_{\rm eq}$ 
is divergent. 
%Here $z$ is the dynamical universal exponent and for QCD critical point $z\approx 3$~\cite{Son:2004iv}. 
%Therefore the non-equilibrium effect would inevitably happen due to the divergence of  $\tau_{\rm eff}$ near a critical point. 
As a result, real time critical fluctuations of QCD matter created in heavy-ion collisions could be significantly different from the corresponding equilibrium values~\cite{Berdnikov:1999ph}. 
%Previously in Ref.~\cite{},
%the growth of the correlation length is found to be limited by this finite time effects. 

%We wish to understand these non-equilibrium effects. 
%How would finite time effects affect the evolution of non-Gaussian cumulants?
In Ref.~\cite{Mukherjee:2015swa},  we studied the real time evolution of non-Gaussian cumulants in the QCD critical regime and implications thereof on searches for the QCD critical point; we will summarize the principal results here.

\section{Evolution equations}
\label{sec:evolution}

%We begin by considering the probability distribution of the critical mode $P(\sigma;\tau)$.  
%Here $\tau$ is the proper time along a given trajectory passing the critical regime. 
%$\sigma$ is the spatially averaged critical field. 
%We note for QCD critical point, 
%the critical field is a linear combination of the chiral condensate and the baryon density~\cite{}. 
%$P(\sigma;\tau)$ encodes full the information on the fluctuations of critical modes.
%Since critical field $\sigma$ is coupled the net baryon number density. 
%The contributions to the moments of net baryon number fluctuations are expected to be proportional to the corresponding moments of the critical field itself
% 
Specifically, in \cite{Mukherjee:2015swa}, we derived a set of evolution equations for the cumulants of critical modes.
Our approach is schematically summarized below:
\begin{eqnarray}
\partial_{\tau}P(\sigma;\tau)= \frac{1}{\tau_{\rm eff}}\hat{{\cal L}}(r,h)P(\sigma;\tau)\,\,\,\, &\Longrightarrow&\,\,\,\,
\frac{d\kappa_{n}(\tau)}{d\tau} = \frac{1}{\tau_{\rm eff}} L_{n}\left[\kappa_{1},\kappa_{2},\ldots,\kappa_{\infty};r,h\right]
\label{FP1}
%\nonumber 
\\
\label{FP2}
&\Longrightarrow& \,\,\,\,  
\frac{d\kappa_{n}(\tau)}{d\tau} = \frac{1}{\tau_{\rm eff}} \bar{L}_{n}\left[\kappa_{1},\kappa_{2},\ldots,\kappa_{n};r,h\right]+{\cal O}\left(\frac{\xi_{\rm eq}}{L_{\rm sys}}\right)
\, . 
\end{eqnarray}
We start with a Fokker-Planck equation on the L.H.S of Eq.~(\ref{FP1}), which describes the relaxation of $P(\sigma;\tau)$ towards equilibrium.
Here  $P(\sigma;\tau)$ is the probability distribution of the (space averaged) critical mode $\sigma$. The proper time $\tau$ characterizes a  trajectory in the ``cross-over" vicinity of the critical regime. 
$\hat{\cal L}$ denotes the differential operator acting on $P(\sigma,\tau)$. 
The relaxation time $\tau_{\rm eff}\sim \xi^{z}_{\rm eq}$, where $z$ is the dynamical universal exponent; for the QCD critical point, $z\approx 3$~\cite{Son:2004iv}. 
%Therefore the non-equilibrium effect would inevitably happen due to the divergence of  $\tau_{\rm eff}$ near a critical point. 
 %is controlled by $\tau_{\rm eff}$ and scales as $\xi^{3}_{\rm eq}$ in the critical regime.
One thereby implements dynamical universality.  Since we are interested in the evolutions of cumulants, 
we translate the Fokker-Planck equation into an (infinite) set of evolution equations for cumulants $\kappa_{n}(\tau), n=1,2,3, \ldots$, see R.H.S of (\ref{FP1}). 
$L_{n}[\kappa_{1},\kappa_{2},\ldots,\kappa_{\infty};r,h]$ is a polynomial of $\kappa_{n}$ with coefficients depend on $r,h$. 
It is easy to convince oneself that this set of equations is exact and contains the same amount of information as the original Fokker-Planck equation.  However, its practical application is limited since the evolution of the lower cumulants depends on that of the higher cumulants. 
Our key observation is that in the regime that the equilibrium correlation length $\xi_{\rm eq}$ is much smaller than the size of the system $L_{\rm sys}$ (but still larger than any other microscopic scale $l_{\rm mic}$),
the evolution of lower cumulants are effectively decoupled from those of higher cumulants. 
We thus arrive at (\ref{FP2}), describing the closed form evolution cumulants~\cite{Mukherjee:2015swa}.

\section{Results}
\subsection{Universal and non-universal inputs}
We wish to apply (\ref{FP2}) to study the evolution of cumulants in QCD critical regime. 
The results depend on both universal and non-universal inputs. 
Universal inputs include the dependence of the equilibrium distribution on the well known Ising variables $r$ and $h$ and the previously noted scaling form of  $\tau_{\rm eff}\sim \xi^{3}_{\rm eq}$. 
 
We also need to specify inputs to the evolution that are non-universal. In particular, 
the mapping between the QCD variables $T,\mu$ and $r,h$ is non-universal. 
We will use the conventional linear mapping relation, i.e. $(T-T_{c})\sim h, (\mu-\mu_{c})\sim r $ where $(T_{c},\mu_{c})$ denotes the location of the critical point in QCD phase diagram.
%The trajectory passing the critical regime in principle can be determined by full hydrodynamic simulations. 
We take simplified yet phenomenologically motivated 
%parameterizations
parametrizations for trajectories passing the critical regime. 
%
%================= vDec. 22th =======================
%
%These include $\tau_{\rm rel}$, which is $\tau_{\rm eff}$ at the boundary of the critical regime and $\tau_{I}$, the proper time at which the system enters into the critical regime. The latter is roughly proportional to the duration of the system in the critical regime. After fixing these parameters, our result will only depend on the dimensionless ratio $\tau_{\rm rel}/\tau_{I}$. 
%
%================= vDec. 22th =======================
These include $\tau_{I}$, the proper time at which the system enters into the critical regime. 
We note the duration of the system in the critical regime  is roughly proportional to $\tau_{I}$. 
We also introduced $\tau_{\rm rel}$, which is $\tau_{\rm eff}$ at the boundary of the critical regime.
After fixing these parameters, our result will only depend on the dimensionless ratio $\tau_{\rm rel}/\tau_{I}$. 
In other words, the evolution of critical fluctuations will depend on the ratio between the magnitude of the relaxation time and the time it spends in the critical regime.

\subsection{Evolution along a representative trajectory}

\begin{figure}[hbt!]
\begin{center}
\includegraphics[width=0.31\textwidth,height=0.22\textwidth]{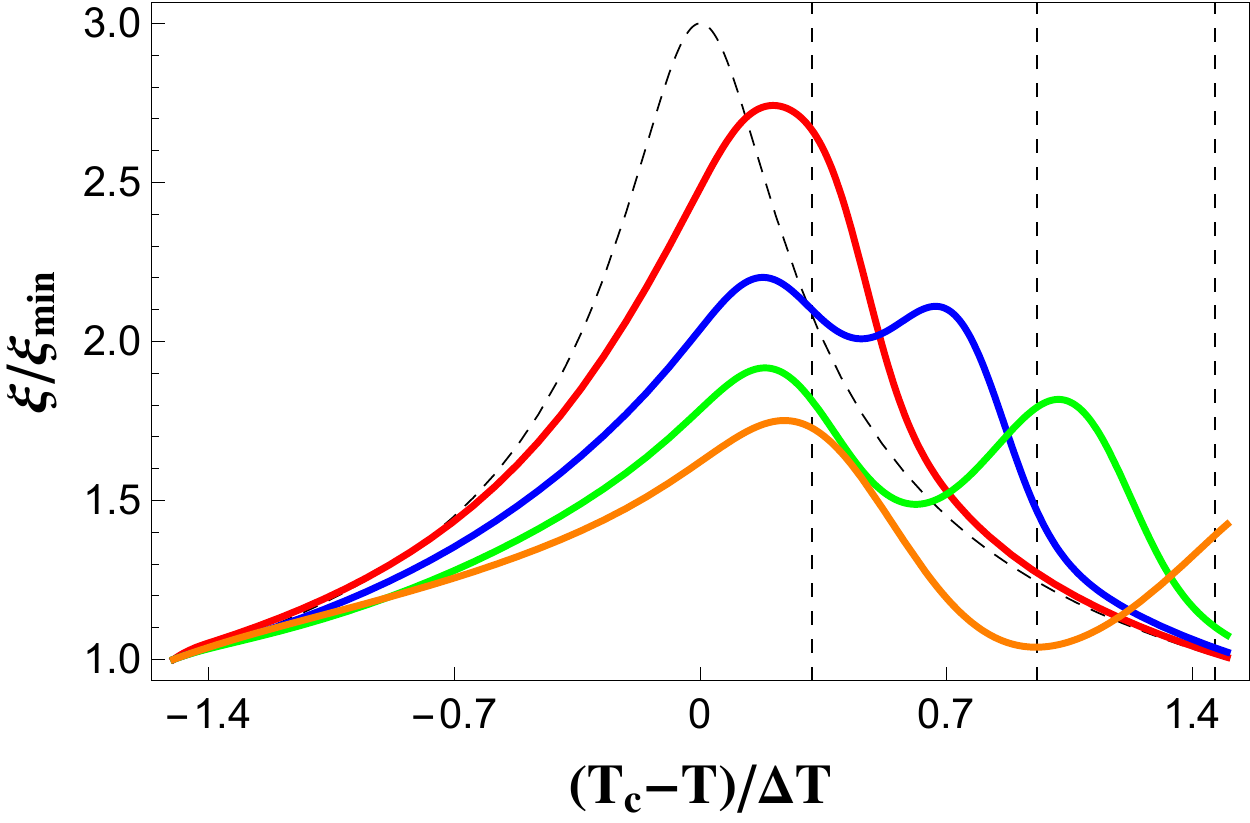}
\includegraphics[width=0.31\textwidth,height=0.22\textwidth]{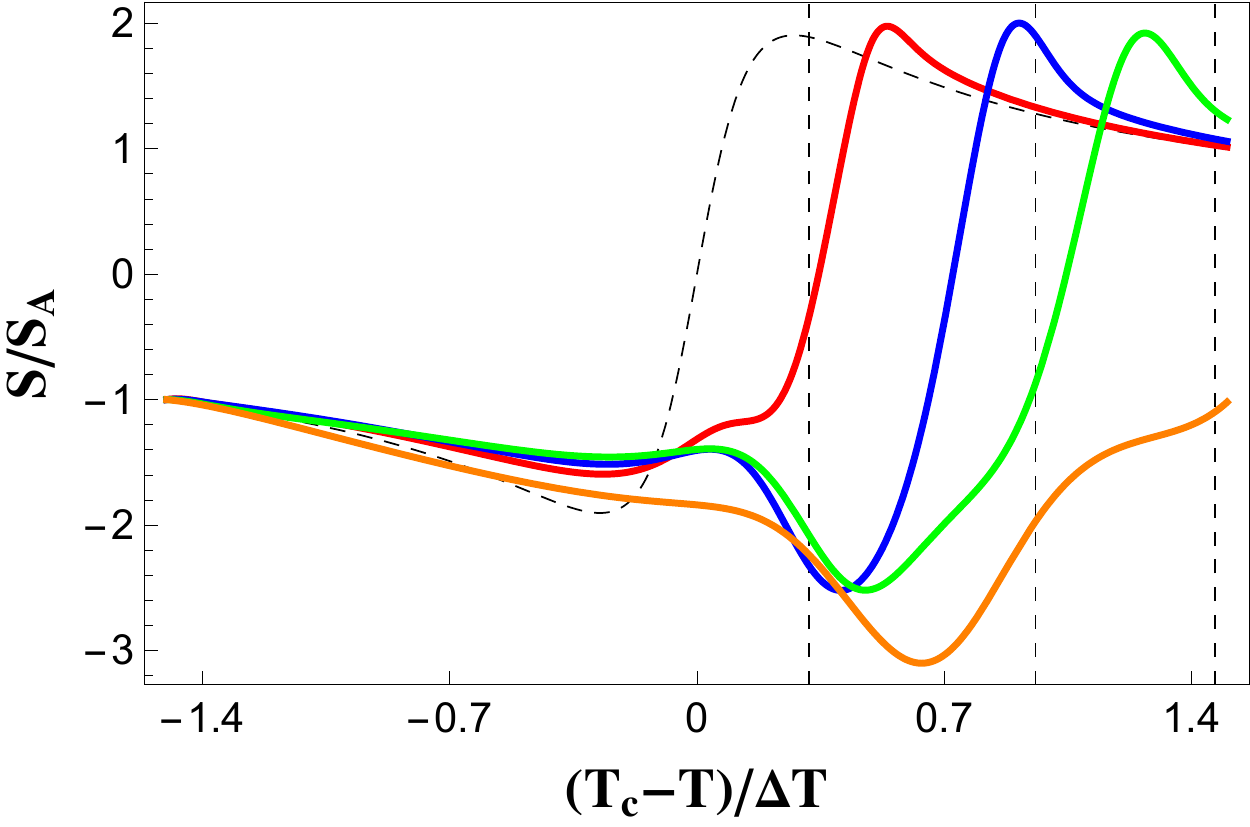}
\includegraphics[width=0.31\textwidth,height=0.22\textwidth]{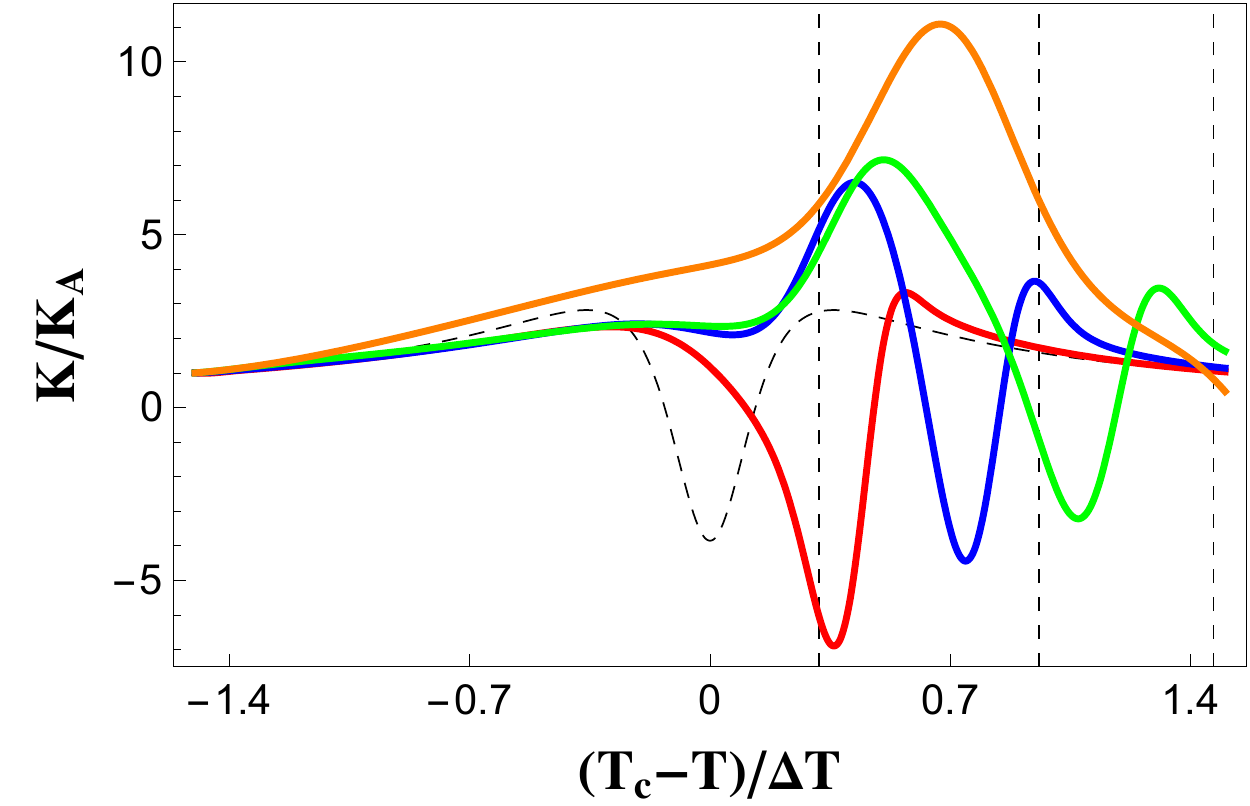}
\caption{
\label{fig:trajA}
(Color online)
The evolution of non-equilibrium critical fluctuations along a representative trajectory. 
Results for 
$\tau_{\rm rel}/\tau_I=0.005,0.02,0.05,0.2$ are shown in red, blue, green, orange curves respectively. 
The dashed curves plot the corresponding equilibrium values. 
All results are normalized by the corresponding equilibrium value at
the end point of trajectory A .
The dashed vertical lines illustrate the $T$ at the point where the
a freeze-out curve intersects with trajectory A (corresponding to freeze-out curves of type I, II, III
respectively).
See also~Fig.~\ref{fig:BES} (left). 
%\vspace{-0.05in}
\label{fig_data}
}
\end{center}
\end{figure}

We now present our results. 
It is instructive to first study the evolution of cumulants along a representative trajectory, trajectory A as labelled in Fig.~\ref{fig:trajA}. 
We introduce a ``non-equilibrium correlation length'' $\xi\equiv \sqrt{\kappa_{2}T/V}$ where $V$ is the volume and $T$ is the temperature, skewness $S=\kappa_{3}/\kappa^{2/3}_{2}$  and kurtosis $\kappa_{4}/\kappa^{2}_{2}$. 
Their evolutions with different $\tau_{\rm rel}/\tau_{I}$ are summarized in Fig.~\ref{fig:trajA}. 

Let us take a look at the impact of critical slowing down on the non-equilibrium correlation length $\xi$. 
%There are curses and blessings out of the critical slowing down. 
On the one hand, the effects of critical slowing down delay the growth of the $\xi$.  On the other, the same effects also slow the decay of $\xi$.
For example, away from $T_{c}$,
the non-equilibrium value of $\xi$ can be significantly larger than the equilibrium value.
The system remembers it was near the critical point. 

Turning to the non-equilibrium evolution of skewness and kurtosis,  one observes immediately that they do not necessarily follow the evolution of the corresponding equilibrium cumulants. The differences in both the magnitude and sign are remarkable .
This is because the evolution of $S$ and  $K$ will not only depend on their deviation from the equilibrium value, but also on the non-equilibrium values of other cumulants.

\subsection{Mimicking the Beam energy scan}

\begin{figure}[hbt!]
\begin{center}
\includegraphics[width=0.3\textwidth,height=0.24\textwidth]{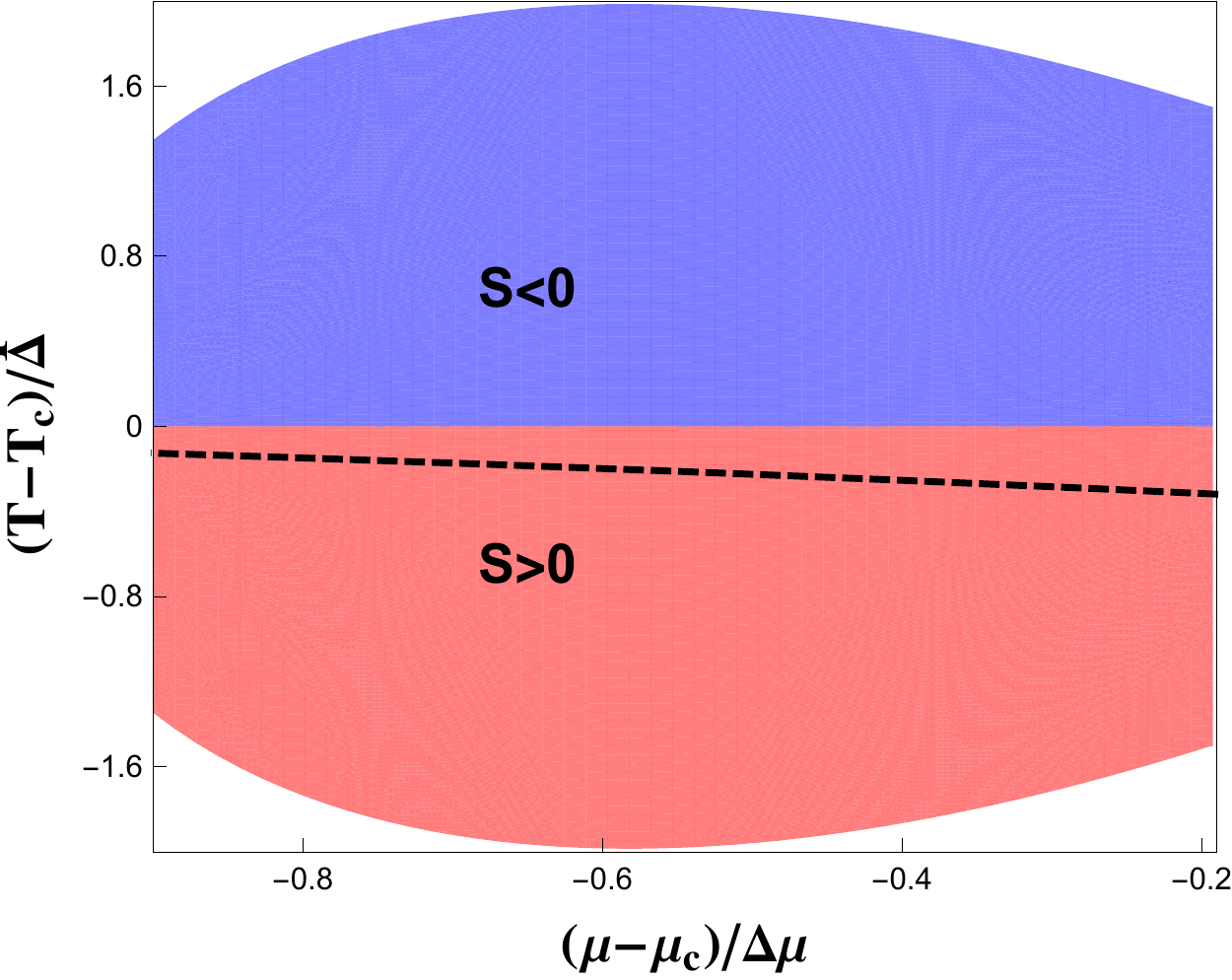}
\includegraphics[width=0.31\textwidth,height=0.24\textwidth]{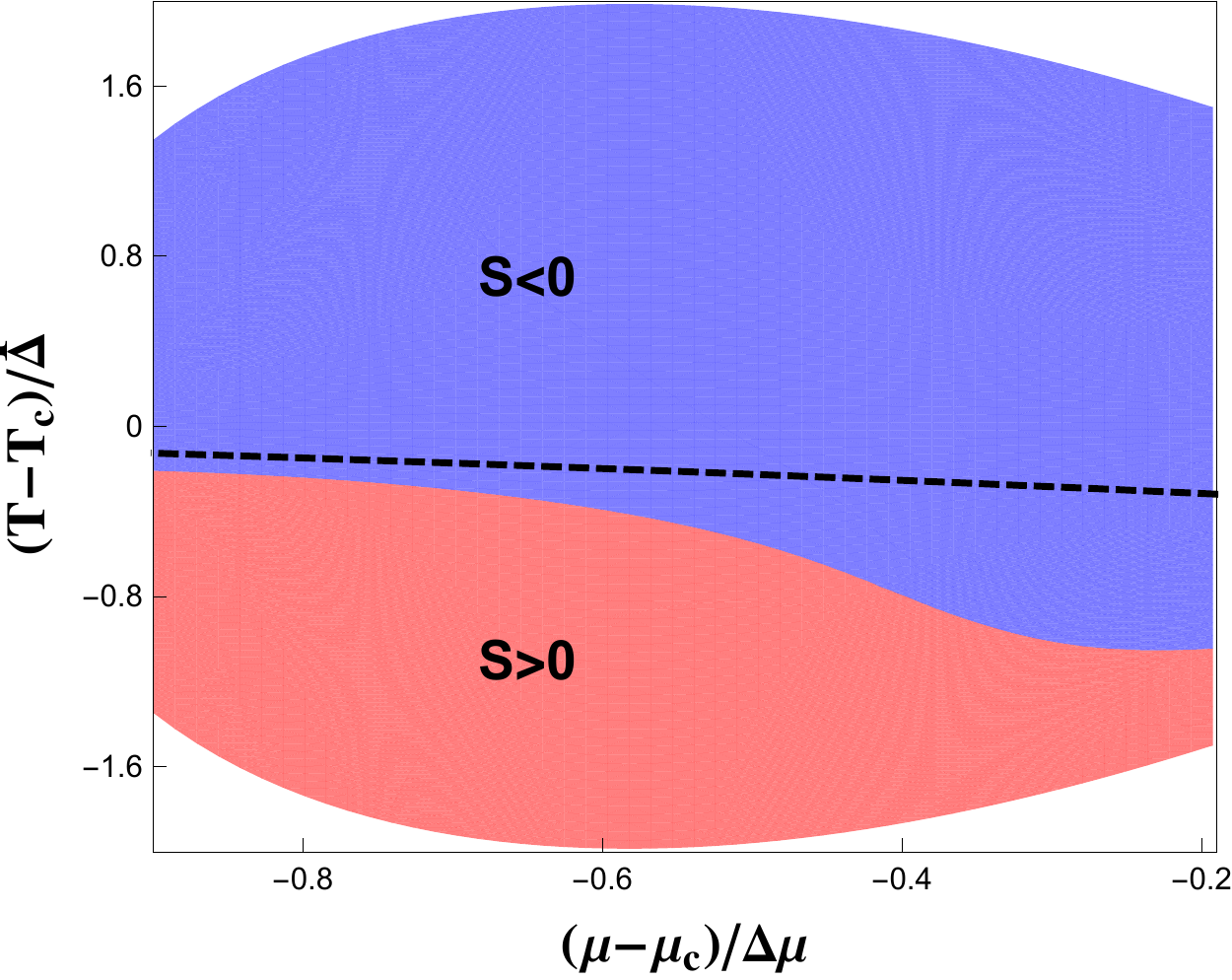}
\includegraphics[width=0.31\textwidth,height=0.24\textwidth]{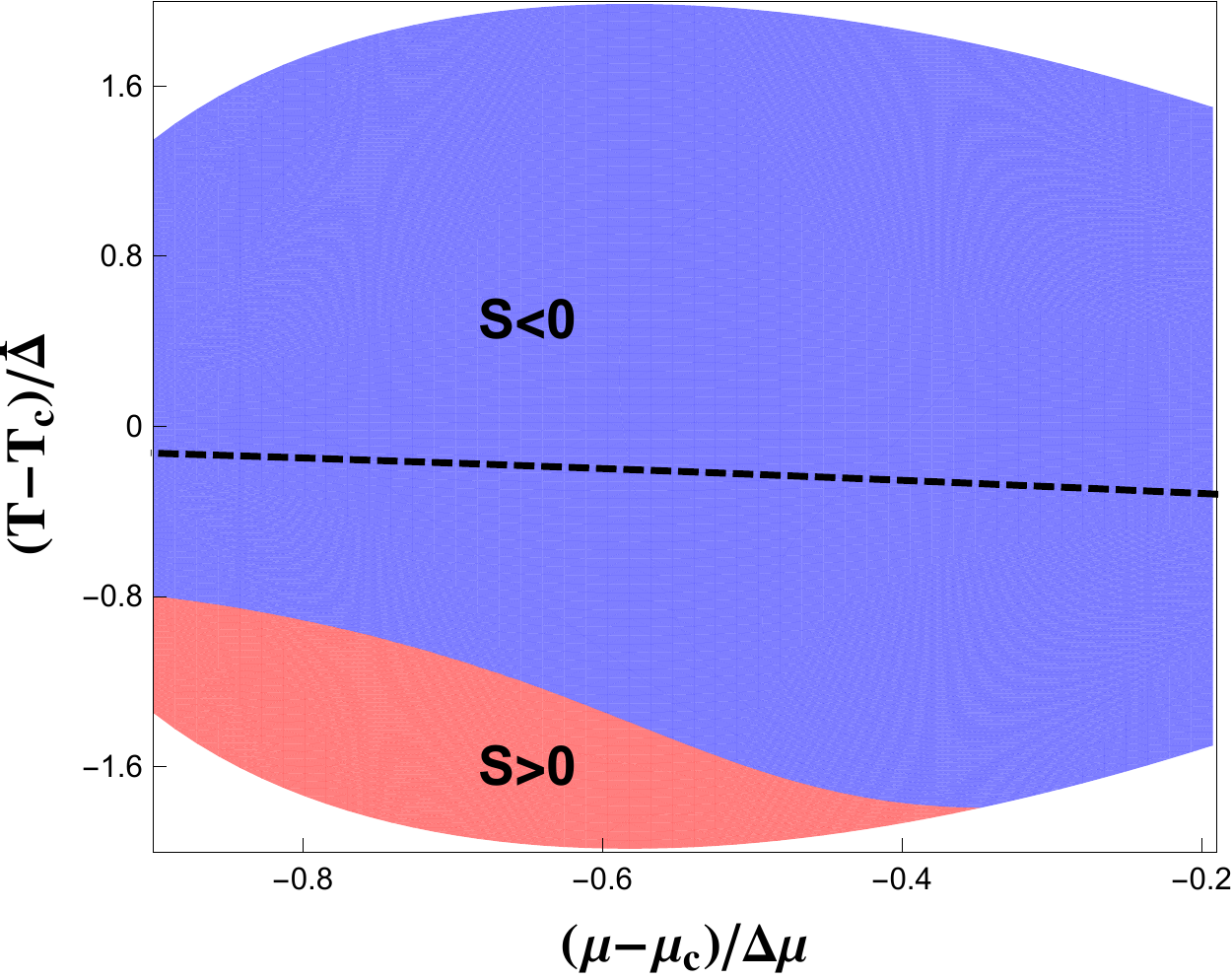}
\includegraphics[width=0.3\textwidth,height=0.24\textwidth]{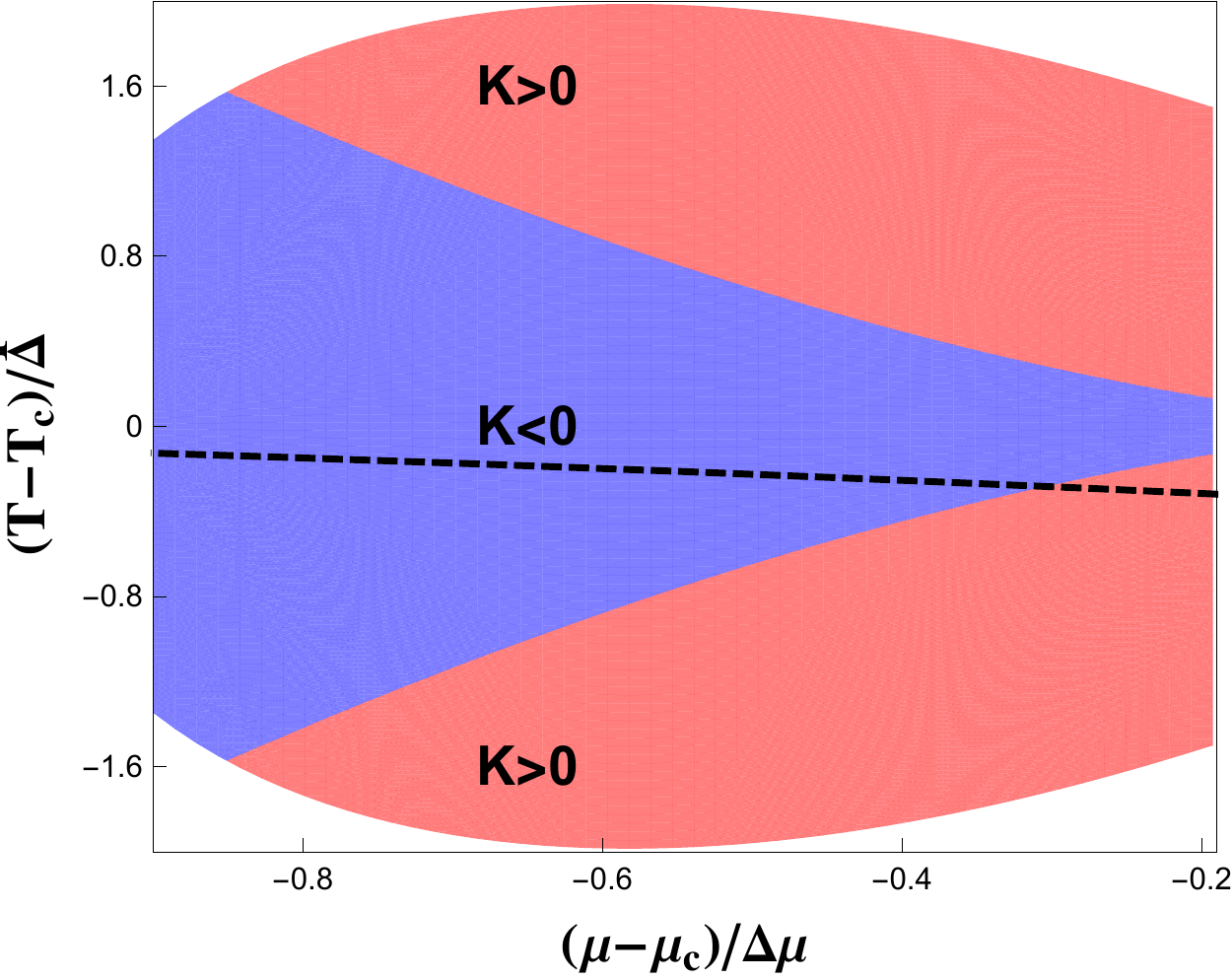}
\includegraphics[width=0.31\textwidth,height=0.24\textwidth]{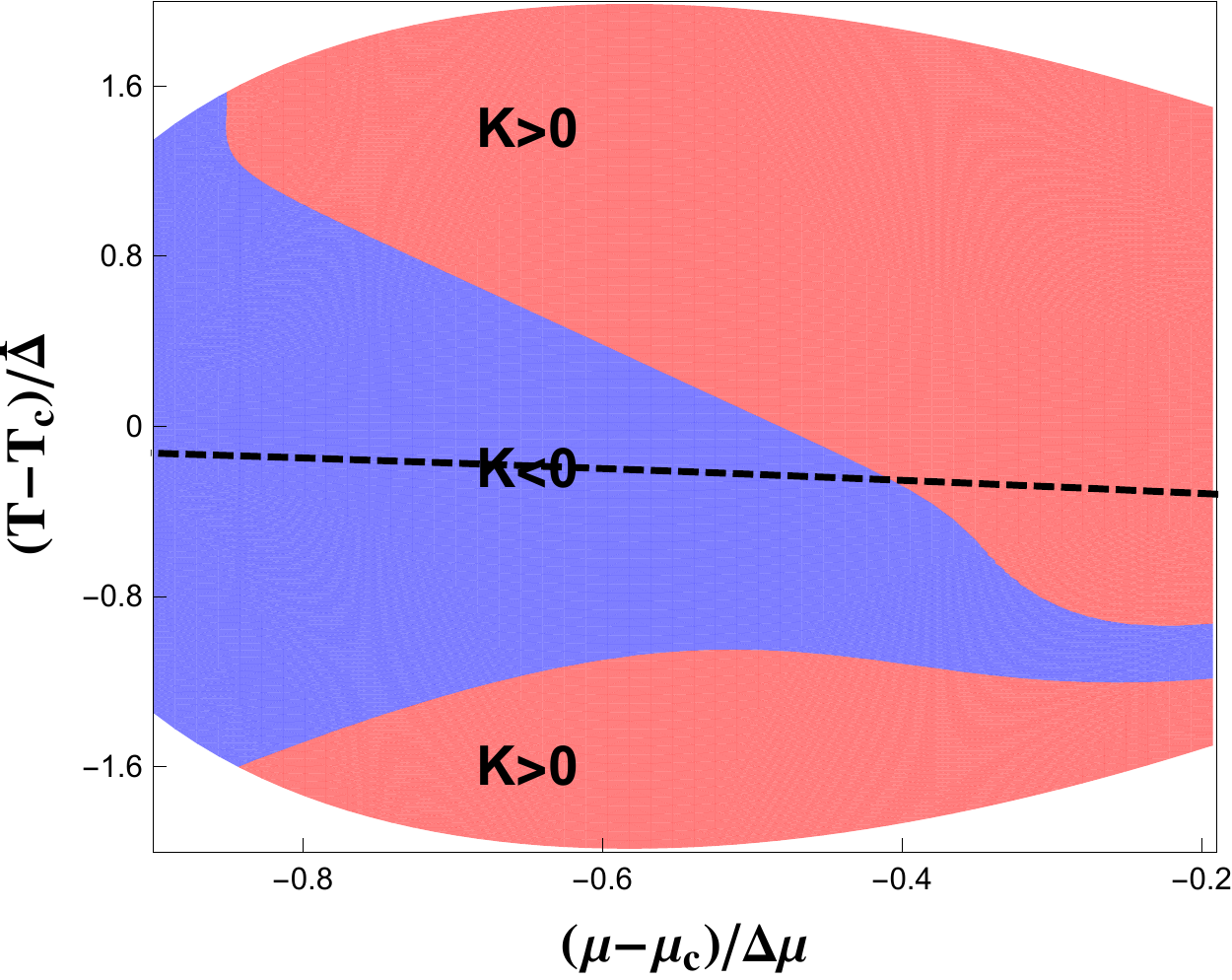}
\includegraphics[width=0.31\textwidth,height=0.24\textwidth]{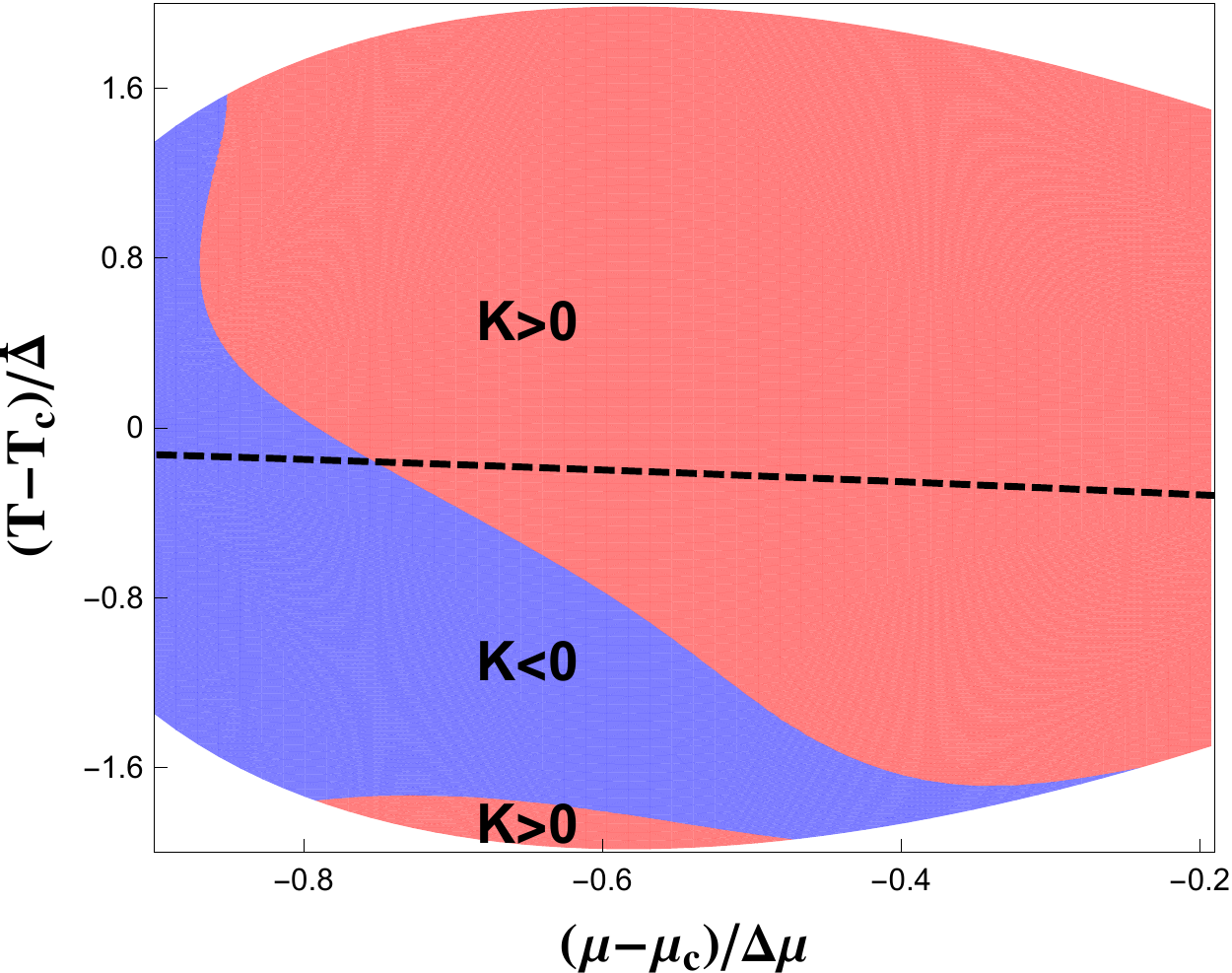}
\caption{(Color online)
%\vspace{-0.05in}
\label{fig:SandK}
Contour plot of equilibrium (left) and non-equilibrium non-Gaussian cumulants with $\tau_{\rm rel}/\tau_{I}=0.05$ (middle) and $\tau_{\rm rel}/\tau_{I}=0.2$ (right). The region of positive and negative non-Gaussian cumulants are shown in red and blue respectively.
The dash curve schematically illustrates the location of the freeze-out curve. 
(Top panel) The skewness $S$. (Bottom panel) The kurtosis $K$. 
}
\end{center}
\end{figure}

We shall now discuss the implications of non-equilibrium effects for the ``Beam energy scan''  (BES) program. 
To mimic the BES, we solved Eq.~(\ref{FP2}) for all trajectories broadly spanning the critical regime. We will first focus on the signs of the skewness $S$ and kurtosis $K$;  close to equilibrium, 
we expect that the critical contribution to the skewness is positive and that kurtosis will flip sign when the critical regime is scanned from the cross-over side of the critical regime. In Fig.~\ref{fig:SandK} (left), we demonstrate what our expectations are for these equilibrium values on the cross-over side of the critical regime. 

What do the non-equilibrium skewness $S$ and kurtosis $K$ look like? 
Fig.~\ref{fig:SandK} demonstrates the deformation of the boundary where the skewness changes sign.
With increasing $\tau_{\rm real}/\tau_{I}$, the non-equilibrium skewness $S$ becomes negative in a larger portion of the area below the cross-over line. This is a manifestation of the remembrance of things past: the system was passing the regime in which the equilibrium skewness is negative. 
%the non-equilibrium cumulants carry more information at early times when ; a larger relaxation time would give more weight to early time contributions.
A similar deformation of the boundary is seen for the kurtosis. Despite this, 
the non-equilibrium cumulants still feature a sign change from positive to negative along a given freeze-out curve as baryon chemical potential $\mu$ is increased (or the beam energy is lowered).
%
%\begin{figure}[hbt!]
%\begin{center}
%\includegraphics[width=0.3\textwidth,height=0.24\textwidth]{K0.pdf}
%\includegraphics[width=0.31\textwidth,height=0.24\textwidth]{K1.pdf}
%\includegraphics[width=0.31\textwidth,height=0.24\textwidth]{K2.pdf}
%\caption{(Color online)
%Kurtosis
%%\vspace{-0.05in}
%\label{fig:K}
%Contour plot of equilibrium (top) and non-equilibrium kurtosis (K) with ?/?I = 0.05 (middle) and ?/?I = 0.02 (bottom). The K > 0 region is shown in red and K < 0 region is shown in blue.
%}
%\end{center}
%\end{figure}

\begin{figure}[hbt!]
\begin{center}
\includegraphics[width=0.3\textwidth,height=0.24\textwidth]{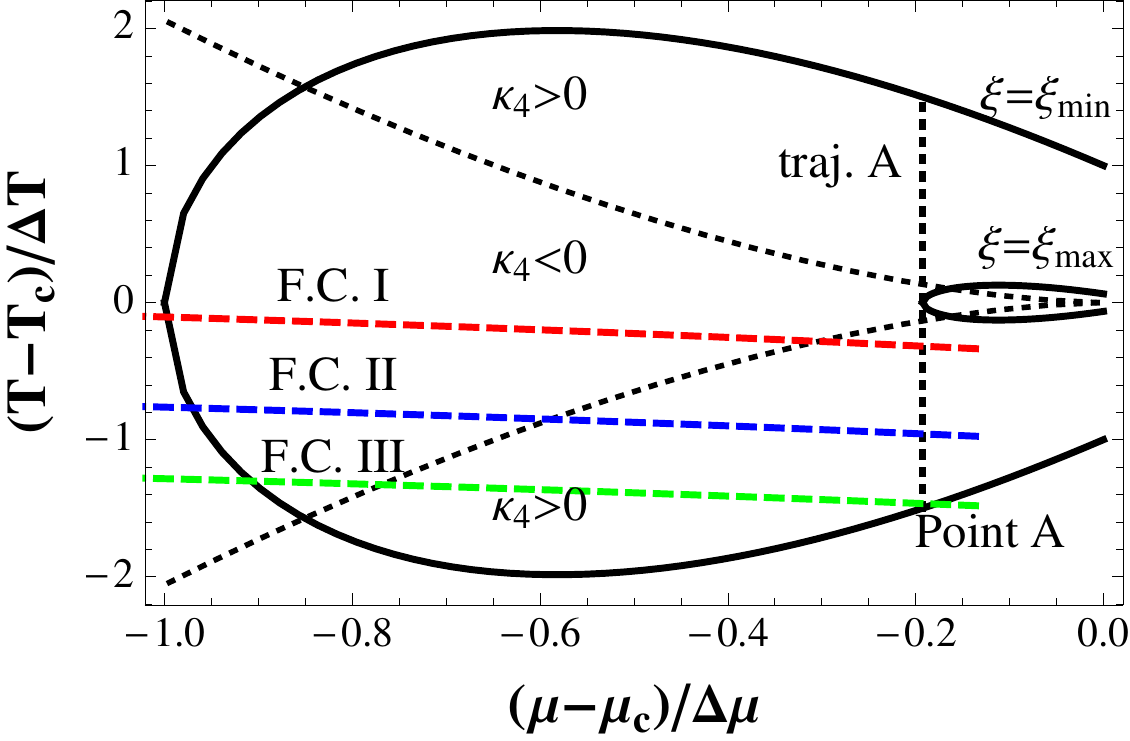}
\includegraphics[width=0.31\textwidth,height=0.24\textwidth]{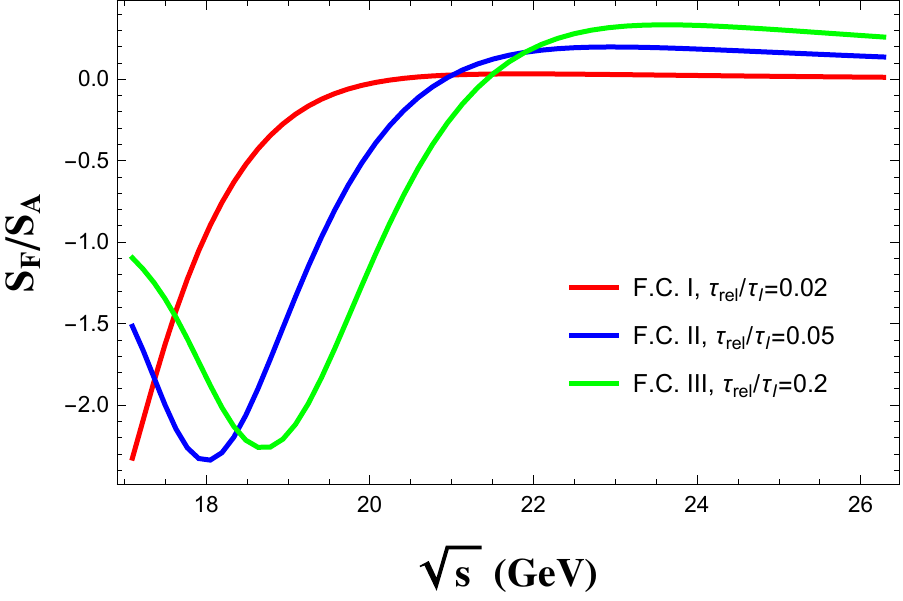}
\includegraphics[width=0.31\textwidth,height=0.24\textwidth]{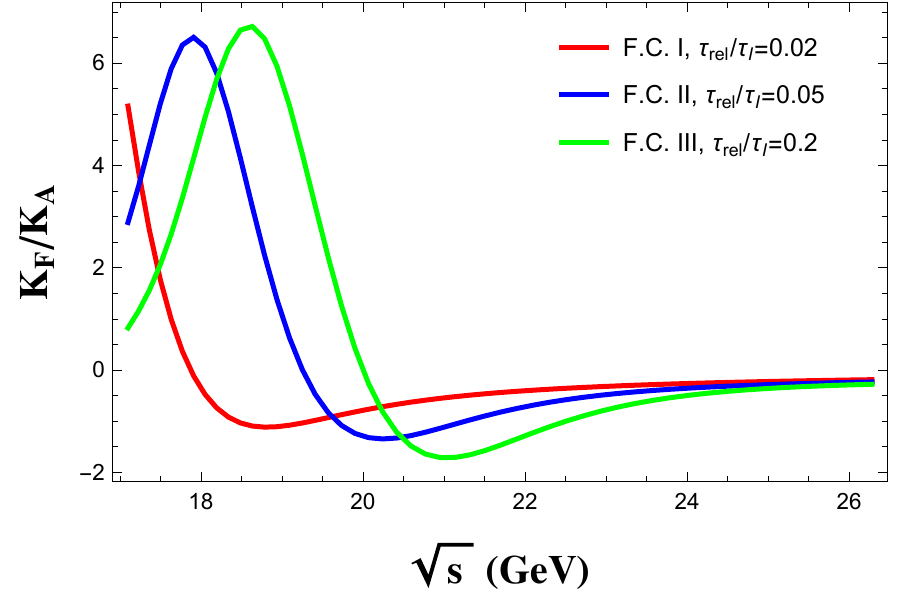}
\caption{(Color online)
%\vspace{-0.05in}
\label{fig:BES}
Schematic illustration of possible $\sqrt{s}$-dependence of non-equilibrium non-Gaussian cumulants on the
freeze-curves with different choices of the relative position of
freeze-out curves and $\tau_{\rm rel}$.
Left: Sketch of the cross-over side of the critical regime. 
Different freeze-out curves (F.C.) are shown in red (upper ), blue (middle) and green (lower) dashed curves, corresponding to type I, II, III respectively. Middle: the non-equilibrium skewness $S$ vs $\sqrt{s}$. Right: the non-equilibrium skewness $K$ vs $\sqrt{s}$. 
%They are neither fittings nor predictions.
}
\end{center}
\end{figure}

Finally, we explore the behavior of cumulants as a function of $\sqrt{s}$ as the QCD phase diagram is scanned.
For this purpose, we convert the $\mu$ dependence into the $\sqrt{s}$ dependence. We must interpret the results presented in Fig.~\ref{fig:BES} with caution. We do not aim to quantitatively fit data or make detailed predictions for future experiments. Our purpose is to illustrate the complications introduced by non-equilibrium effects in interpreting data. In particular, we show that very similar curves, as a function of $\sqrt{s}$,  can be obtained by different combinations of freeze-out curves and relaxation times.

\section{Summary}
In summary, the most pressing issue in the current search for the QCD critical point are  the presence of incontrovertible non-equilibrium effects. 
We derived a coupled set of equations that describe the non-equilibrium evolution of cumulants of critical fluctuations.
We demonstrated that skewness and kurtosis can differ significantly in magnitude as well as in sign from equilibrium expectations. 
We examined the implications of our study for the critical point search in heavy-ion collisions

\section*{Acknowledgements}
This work was supported by DOE Grant No. DE- SC0012704. 
RV thanks the Institut f\"{u}r Theoretische Physik, Heidelberg for their kind hospitality and the Excellence Initiative of Heidelberg University for support.



%% References


\bibliographystyle{elsarticle-num}



\end{document}